\documentclass[aps]{revtex4-2}

\usepackage{amsmath,amssymb,amsfonts}
\usepackage{graphicx}
\usepackage{bm}
\usepackage{physics}
\usepackage{hyperref}
\usepackage{xcolor}
\usepackage{placeins}

\begin{document}

\title{Axion misalignment as a synchronization phenomenon}

\author{Veronica Sanz}
\affiliation{Instituto de F\'isica Corpuscular (IFIC), Universidad de Valencia-CSIC, E-46980 Valencia, Spain}

\begin{abstract}
We propose a dynamical reinterpretation of axion misalignment as an emergent collective phenomenon.
Drawing an explicit parallel between axion field dynamics and synchronization in coupled oscillator systems, we show that a macroscopic axion phase can arise dynamically from initially incoherent configurations through gradient-driven ordering in an expanding Universe.
In this framework, the misalignment angle is not a fundamental initial condition but a collective variable that becomes well defined only once phase coherence develops.
Using a three-dimensional lattice study with full second-order dynamics, we show that this collective phase is selected prior to the onset of axion oscillations in the regimes studied, providing dynamical support for the standard misalignment picture.
This perspective offers a new way of organizing axion initial-condition sensitivity and reframes small-angle assumptions in terms of the efficiency of phase ordering in the early Universe.
\end{abstract}

\maketitle

\section{Introduction}

The axion remains one of the most compelling candidates for cold dark matter, arising
naturally from the Peccei--Quinn (PQ) solution to the strong $CP$ problem
\cite{PecceiQuinn1977,Weinberg1978,Wilczek1978}.
In the standard cosmological picture, the axion relic abundance is generated by the
misalignment mechanism, in which the axion field begins coherent oscillations when
the Hubble rate drops below the axion mass,
$3H(T_{\rm osc})\simeq m_a(T_{\rm osc})$
\cite{PreskillWiseWilczek1983,AbbottSikivie1983,DineFischler1983}.
Assuming a spatially homogeneous initial field value $\theta_i$, the present-day
abundance scales as
$\Omega_a \propto f_a^{\,p}\theta_i^2$, up to known corrections
\cite{Turner1986,Marsh2016}.

For large axion decay constants $f_a$, reproducing the observed dark matter density
requires $\theta_i\ll1$, a fact often interpreted in terms of anthropic selection
\cite{Turner1986}.
This reasoning treats the initial misalignment angle as a fundamental environmental
parameter, implicitly assuming that a unique, homogeneous axion phase exists prior
to the onset of oscillations.
However, the axion is a compact phase field whose cosmological evolution is governed
by local dynamics, and the existence of a single global angle is itself a dynamical
question.

In this work we point out that the axion field equations describe a system that is
mathematically equivalent to a network of locally coupled, damped phase variables
with time-dependent coupling set by cosmological expansion.
Such systems are known to exhibit synchronization and phase-ordering phenomena, in
which macroscopic collective variables emerge dynamically from initially incoherent
microscopic phases.
Synchronization was originally introduced by Kuramoto in the context of coupled
oscillators \cite{Kuramoto1975,Kuramoto1984} and has since been recognized as a
universal mechanism underlying collective phase coherence in a wide range of physical
systems \cite{Acebron2005}. In spatially extended systems with local coupling, synchronization and phase ordering describe the same dynamical process viewed from microscopic and macroscopic perspectives, respectively.

Adopting this perspective, the axion misalignment angle is naturally identified with
the collective phase of the axion field, which becomes well defined only once
dynamical coherence has been established.
The appropriate macroscopic variables are the synchronization order parameter
$R$ and the associated collective phase $\Psi$; a single ``initial angle'' exists
only in the regime $R\simeq1$.
The standard misalignment formula is therefore valid only after the axion field has
undergone a phase-ordering process driven by gradient interactions, Hubble damping,
and the time-dependent axion potential.

This reframing does not modify the axion equations of motion, nor does it alter the standard relic abundance calculation within its usual domain of applicability.
Rather, it makes explicit the dynamical assumptions implicit in the use of a homogeneous initial angle, namely that sufficient phase ordering has occurred prior to the onset of oscillations.
From this perspective, anthropic small-angle arguments are reinterpreted not as statements about fundamental initial data, but as assumptions about the efficiency of collective ordering in the early Universe.

We emphasize that the synchronization perspective adopted here differs from previous uses of related language in axion physics.
Synchronization has been discussed in the context of axion--Josephson or axion--superconductor analogies, where an axion phase couples to an external coherent oscillator~\cite{Beck2013}, as well as in defect dynamics, where ``phase locking'' refers to the attachment of strings and domain walls once the axion potential turns on~\cite{ChadhaDay2014}.
By contrast, we focus on synchronization among axion field degrees of freedom themselves, as a cosmological phase-ordering process that determines when a macroscopic axion angle exists and how its value is selected.
Similarly, while the standard distinction between Peccei--Quinn symmetry breaking before or after inflation is usually framed in terms of horizon size and defect formation~\cite{Marsh2016}, we reinterpret it here in terms of the efficiency and scale of collective phase ordering.

In the following sections, we make the equivalence between axion dynamics and
synchronization explicit, introduce the relevant collective variables, and test their emergence with three-dimensional numerical simulations incorporating cosmological inputs.

\section{Axion field dynamics as a coupled phase system}

\subsection{Continuum axion equation in an expanding Universe}

After Peccei--Quinn symmetry breaking, the axion field can be written as
$\phi(x,t)=f_a\,\theta(x,t)$, where $\theta$ is a compact phase variable,
$\theta\equiv\theta+2\pi$
\cite{Weinberg1978,Wilczek1978}.
In a spatially flat Friedmann--Robertson--Walker background, the axion phase obeys
\begin{equation}
\ddot{\theta}(x,t)
+3H(t)\dot{\theta}(x,t)
-\frac{1}{a^2(t)}\nabla^2\theta(x,t)
+m_a^2(T)\sin\theta(x,t)=0 ,
\label{eq:eom_cont}
\end{equation}
where $H(t)$ is the Hubble rate, $a(t)$ the scale factor, and $m_a(T)$ the
temperature-dependent axion mass
\cite{PreskillWiseWilczek1983,AbbottSikivie1983,DineFischler1983,Marsh2016}.

Equation~\eqref{eq:eom_cont} contains three key ingredients: local phase coupling
through spatial gradients, dissipation via Hubble friction, and a time-dependent
restoring force that becomes effective near the QCD scale.
This structure is characteristic of phase-ordering dynamics in systems with compact
order parameters
\cite{Kibble1976,Bray1994}.

\subsection{Discretization and emergence of a coupling network}

To make the connection explicit, we discretize space on a comoving lattice with spacing
$\Delta x$ and define $\theta_i(t)\equiv\theta(x_i,t)$.
The Laplacian is approximated by nearest-neighbor differences,
\begin{equation}
\nabla^2\theta(x_i)
\;\longrightarrow\;
\frac{1}{(\Delta x)^2}
\sum_{j\in\langle i\rangle}
\big(\theta_j-\theta_i\big) .
\label{eq:lap_disc}
\end{equation}
Substituting into Eq.~\eqref{eq:eom_cont} yields
\begin{equation}
\ddot{\theta}_i
+3H\dot{\theta}_i
+\sum_{j\in\langle i\rangle}
J_{ij}(t)\big(\theta_i-\theta_j\big)
+m_a^2(T)\sin\theta_i
=0 ,
\label{eq:eom_lattice_linear}
\end{equation}
with a time-dependent coupling
\begin{equation}
J_{ij}(t)=\frac{1}{a^2(t)(\Delta x)^2}
\qquad (j\in\langle i\rangle),
\label{eq:Jij}
\end{equation}
and $J_{ij}=0$ otherwise.
The coupling strength therefore decreases with cosmic expansion and is fully fixed by
the background geometry and the spatial resolution; it introduces no new free
parameters.
Discretizations of this form are standard in numerical studies of axion field dynamics
and defect evolution
\cite{Sikivie2008,Marsh2016}.

Since $\theta$ is a compact variable, it is convenient to rewrite the linear coupling
in Eq.~\eqref{eq:eom_lattice_linear} in a periodic form,
\begin{equation}
\sum_{j\in\langle i\rangle}
J_{ij}(t)\big(\theta_i-\theta_j\big)
\;\longrightarrow\;
\sum_{j\in\langle i\rangle}
K_{ij}(t)\sin(\theta_j-\theta_i),
\label{eq:sine_coupling}
\end{equation}
which is equivalent in the regime of small phase differences and respects the
underlying $2\pi$ periodicity.
The effective coupling matrix is therefore
\begin{equation}
K_{ij}(t)=\frac{1}{a^2(t)(\Delta x)^2},
\label{eq:Kij}
\end{equation}
identical for all nearest-neighbor pairs.
This representation makes the analogy with coupled phase systems manifest.

\subsection{Relation to synchronization dynamics}

Equations~\eqref{eq:eom_lattice_linear} and~\eqref{eq:sine_coupling} describe a network
of identical phase variables with local coupling, dissipation, and a time-dependent
on-site potential.
This structure is mathematically equivalent to a second-order generalization of the
Kuramoto model for synchronization, with vanishing intrinsic frequencies and a
time-dependent coupling strength
\cite{Kuramoto1975,Kuramoto1984,Acebron2005}.

In this language, the axion gradient term enforces local phase alignment, while Hubble
friction drives relaxation toward ordered configurations.
The emergence of a homogeneous axion mode is therefore not an assumption but a
dynamical outcome of phase ordering.
Whether and when a single global phase develops depends on the competition between
gradient coupling, cosmological damping, and the turn-on of the axion potential.

\section{Emergent collective variables and the meaning of the misalignment angle}

\subsection{Order parameter and phase coherence}

The axion field equation describes the evolution of a compact phase subject to local
coupling and dissipation. In such systems, the appropriate macroscopic description is
not given by individual phases but by collective variables characterizing coherence.
We therefore introduce the complex order parameter
\begin{equation}
R(t)e^{i\Psi(t)} \equiv \frac{1}{V}\int_V \dd^3x\, e^{i\theta(x,t)} ,
\label{eq:order_param}
\end{equation}
where $R(t)\in[0,1]$ measures the degree of phase coherence and $\Psi(t)$ defines a
collective phase.
This definition is standard in the theory of synchronization and phase ordering
\cite{Kuramoto1975,Bray1994,Acebron2005}.

Two limiting regimes are of particular relevance.
For an incoherent field configuration with random phases, one has $R\simeq0$ and no
global phase is well defined.
Conversely, when the field has developed long-range coherence, $R\simeq1$ and the axion
field is well approximated by a homogeneous mode,
\begin{equation}
\theta(x,t)\simeq\Psi(t)\qquad (R\simeq1).
\label{eq:homog_limit}
\end{equation}
Only in this regime does it become meaningful to describe the system in terms of a
single axion angle.

\subsection{Reduction to the homogeneous mode}

When the field has developed strong coherence and the phase distribution is narrowly concentrated around a single branch, the axion energy density reduces to that of the collective mode. In the small-angle regime, $|\Psi|\ll 1$, one finds
\begin{equation}
\rho_a(t)
\simeq
\frac{f_a^2}{2}\left[\dot{\Psi}^2(t)+m_a^2(T)\Psi^2(t)\right],
\label{eq:rho_collective}
\end{equation}
which coincides with the standard homogeneous-field expression used in misalignment
calculations
\cite{PreskillWiseWilczek1983,AbbottSikivie1983,Marsh2016}.

Outside this coherent small-angle regime, the energy density depends on the full field configuration rather than on a single global phase. A decomposition into a collective contribution plus fluctuations is meaningful only when the phase distribution remains sufficiently narrow that a local unwrapping around the emergent branch is well defined. The standard misalignment formula is therefore justified only after dynamical phase ordering has established a coherent collective mode.

\subsection{Oscillation onset and identification of the misalignment angle}

Axion oscillations begin when the axion mass becomes comparable to the Hubble rate,
\begin{equation}
3H(T_{\rm osc}) \simeq m_a(T_{\rm osc}),
\label{eq:Tosc_def}
\end{equation}
after which the comoving axion number density is approximately conserved.
In the standard treatment, one evaluates the energy density at $T_{\rm osc}$ using a
homogeneous initial angle $\theta_i$.
From the present perspective, this procedure is justified only if the axion field has
already become coherent at that time,
\begin{equation}
R(T_{\rm osc}) \simeq 1.
\label{eq:Rcond}
\end{equation}

When this condition holds, the identification
\begin{equation}
\theta_i \equiv \Psi(T_{\rm osc})
\label{eq:theta_i_ident}
\end{equation}
is well defined, and the standard misalignment abundance follows.
If instead $R(T_{\rm osc})<1$, no unique misalignment angle exists and the axion
abundance depends on the full phase distribution rather than on a single collective
variable.

\subsection{Implications for small-angle assumptions}

The common assumption of a small initial angle $\theta_i\ll1$ can thus be reinterpreted
as a statement about the outcome of the axion phase-ordering dynamics.
Specifically, it presumes that a coherent collective phase $\Psi$ exists and that its
value happens to be small at the onset of oscillations.
Whether this is realized depends on the competition between gradient-induced alignment,
Hubble damping, and the time dependence of the axion potential, rather than on an
arbitrary choice of initial conditions.

This observation does not alter the axion equations of motion nor the standard relic
abundance calculation in the regime $R\simeq1$.
It clarifies, however, that the ``initial misalignment angle'' is not a fundamental
input parameter but an emergent collective variable whose existence and value are set
dynamically.

\section{Three-dimensional robustness study}

To test whether the emergence of a collective axion phase is merely a one-dimensional artifact, we perform a three-dimensional numerical study based on the full second-order dynamics. The purpose of this analysis is not to provide a precision cosmological prediction, but to examine directly whether the qualitative mechanism proposed above survives in the physically relevant case of a spatially extended field with propagating degrees of freedom.

We evolve the axion phase on a three-dimensional periodic lattice according to the discretized version of Eq.~\eqref{eq:eom_cont},
\begin{equation}
\ddot{\theta}_{\mathbf n}
+3H(T)\dot{\theta}_{\mathbf n}
-\frac{1}{a^2(T)}\,\Delta \theta_{\mathbf n}
+m_a^2(T)\sin\theta_{\mathbf n}=0,
\label{eq:eom_3d_num}
\end{equation}
where $\mathbf n$ labels lattice sites and $\Delta$ is the nearest-neighbour discrete Laplacian,
\begin{equation}
\Delta \theta_{\mathbf n}
\equiv
\frac{1}{(\Delta x)^2}
\sum_{\mu=1}^{3}
\left(
\theta_{\mathbf n+\hat\mu}
+\theta_{\mathbf n-\hat\mu}
-2\theta_{\mathbf n}
\right).
\label{eq:lap_num}
\end{equation}
The initial condition is a random phase configuration with $\theta_{\mathbf n}\in(-\pi,\pi]$, drawn independently at each site. We monitor the global order parameter
\begin{equation}
R(T)e^{i\Psi(T)}=\frac{1}{N}\sum_{\mathbf n} e^{i\theta_{\mathbf n}(T)},
\label{eq:order_num}
\end{equation}
and compare the onset of coherence with the conventional oscillation condition $m_a(T_{\rm osc})\simeq 3H(T_{\rm osc})$.

\begin{figure}[t]
  \centering
  \includegraphics[width=0.7\linewidth]{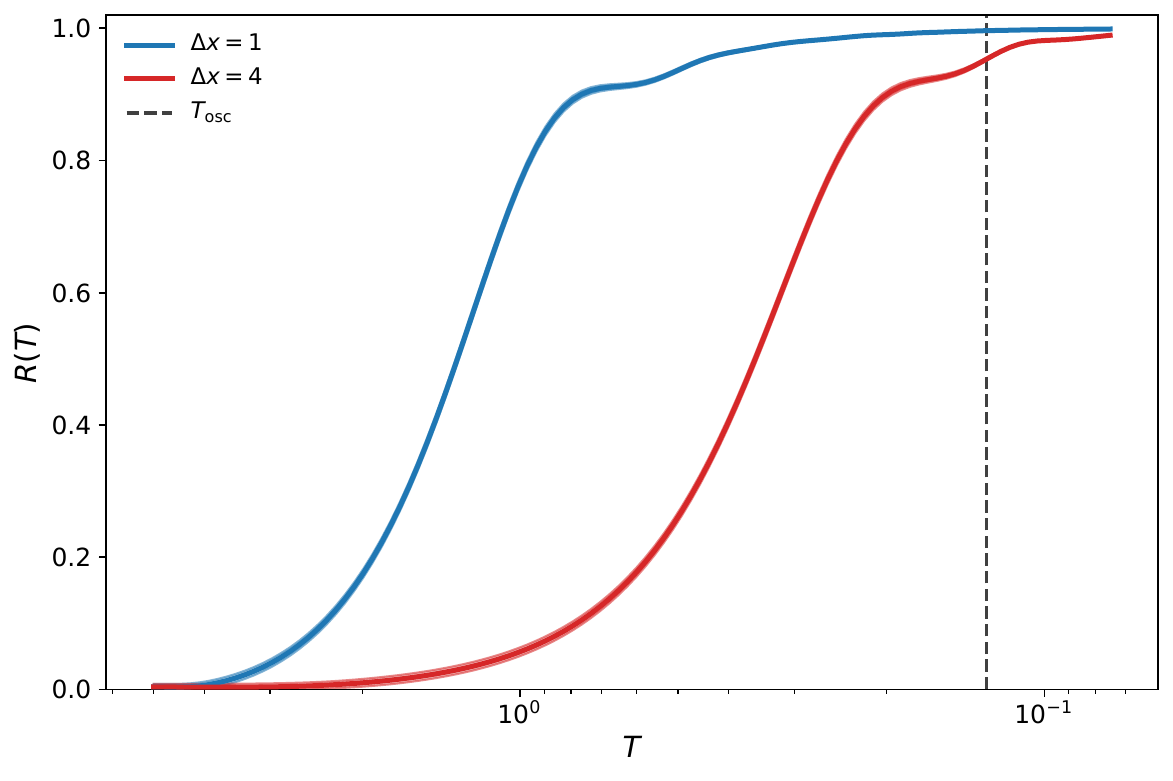}
  \caption{Evolution of the global order parameter $R(T)$ in the three-dimensional full-dynamics study for $\Delta x=1$ and $\Delta x=4$. Solid curves show the mean over all available runs for each $\Delta x$, combining different random seeds and box sizes $L=24,32,40,48$, while the shaded bands indicate the run-to-run spread. The vertical dashed line marks the oscillation temperature $T_{\rm osc}$. Stronger ordering ($\Delta x=1$) leads to earlier and more rapid growth of coherence, whereas weaker ordering ($\Delta x=4$) delays but does not eliminate the onset of collective synchronization.}

  \label{fig:3d_baseline_R}
\end{figure}

Figure~\ref{fig:3d_baseline_R} shows the baseline $32^3$ result. Starting from an incoherent state with $R\simeq0$, the system develops substantial macroscopic coherence during the evolution, and the order parameter approaches unity by the time oscillations begin. This already shows that the emergence of a collective phase is not restricted to a one-dimensional overdamped toy model.

To test robustness, we vary the effective ordering strength by changing the lattice-spacing parameter $\Delta x$, which controls the gradient term in Eq.~\eqref{eq:lap_num}, as a numerical proxy controlling the relative efficiency of gradient-driven ordering in the discretized dynamics. Figure~\ref{fig:3d_dx_scan} shows that weakening gradient-driven ordering delays the onset of coherence and reduces the value of $R(T_{\rm osc})$. The effect therefore persists, but its efficiency is dynamical rather than automatic. In particular, the synchronization perspective is most predictive in the regime where ordering is sufficiently rapid that a well-defined collective phase exists by oscillation onset.

\begin{figure*}[ht!]
  \centering
  \includegraphics[width=0.95\linewidth]{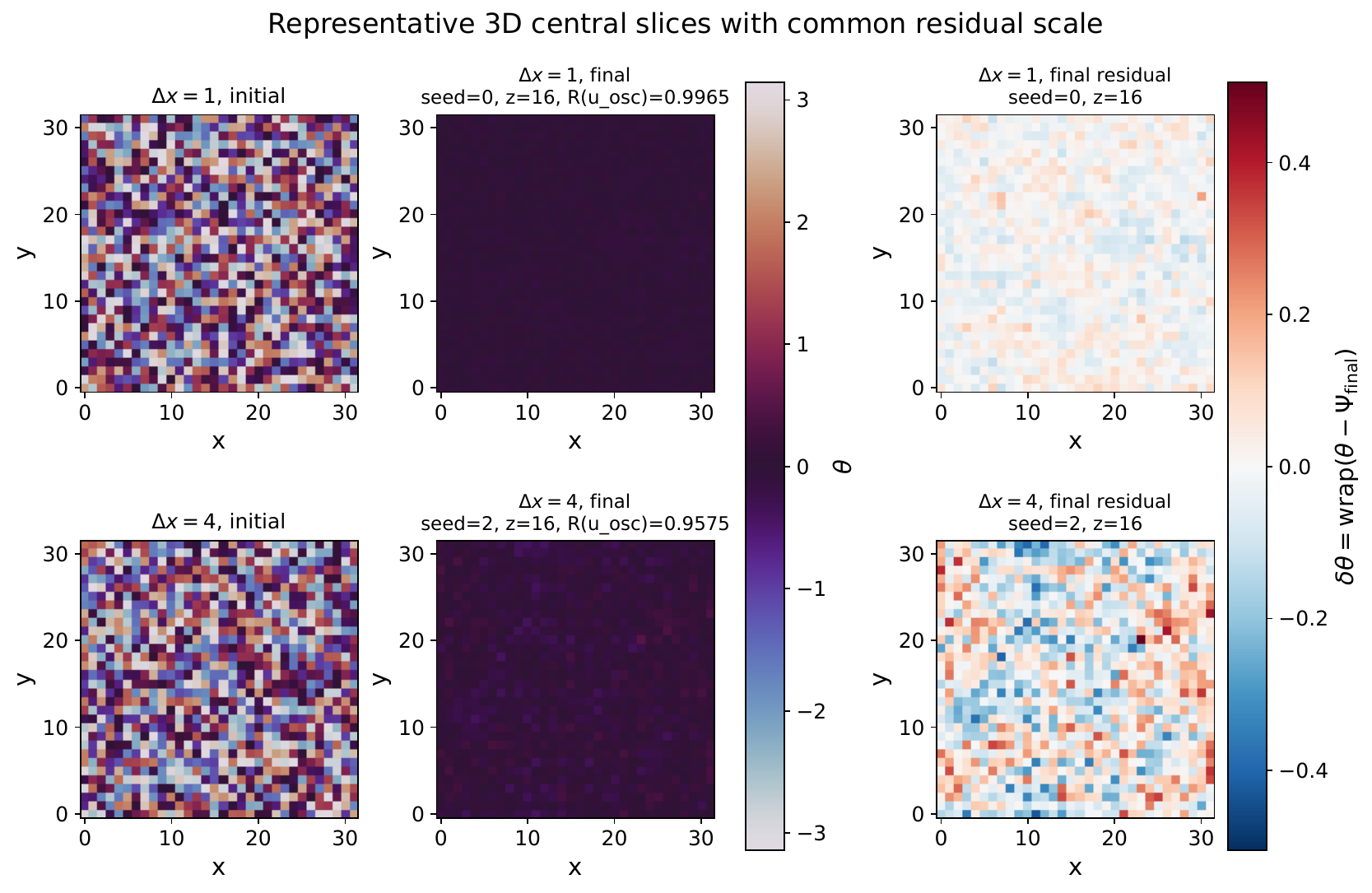}
 
  \caption{Representative central slices of the three-dimensional field for a stronger-ordering case ($\Delta x=1$) and a weaker-ordering case ($\Delta x=4$). In both cases the field evolves from random initial phases to a highly ordered late-time configuration. The residual slice, defined by $\delta\theta=\mathrm{wrap}(\theta-\Psi_{\rm final})$, shows visibly larger residual structure in the weaker-ordering case.}
  \label{fig:3d_slices}
\end{figure*}

Representative central slices of the three-dimensional field are shown in Fig.~\ref{fig:3d_slices}. In both the stronger- and weaker-ordering cases, the field evolves from a random initial configuration to a highly ordered late-time state. The weaker-ordering case retains visibly larger residual fluctuations around the final collective phase, consistent with the lower value of $R(T_{\rm osc})$.

\begin{figure}
    \centering
    \includegraphics[width=0.5\linewidth]{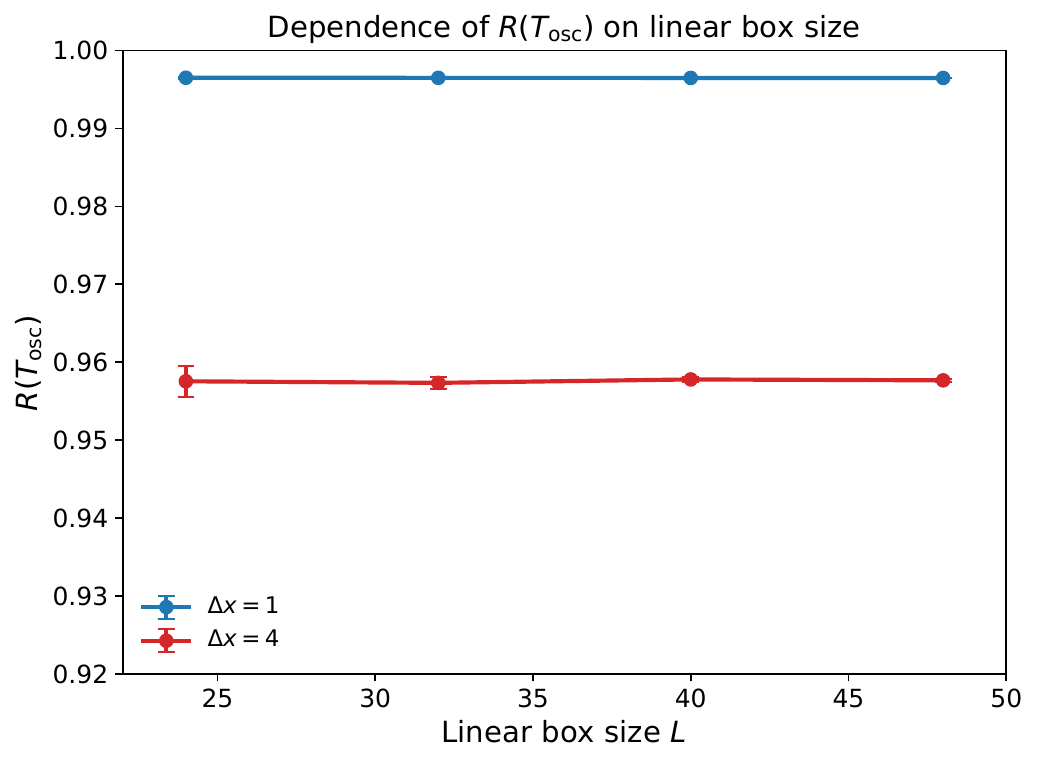}
    \caption{Order parameter as a function of the box size for two representative values of $\Delta x$.}
    \label{fig:3d_size_scan}
\end{figure}

The three-dimensional size check of Fig.~\ref{fig:3d_size_scan} indicates that, over the lattice sizes we were able to explore, the ordering mechanism is not obviously a small-box artifact. While this does not amount to a precision finite-volume analysis, it supports the interpretation of the 3D results as a robustness study rather than as a special feature of a single lattice realization.

These results address directly the main limitation of the original numerical illustration. The qualitative mechanism proposed in this work---namely that the misalignment angle can emerge as a collective variable through phase ordering---survives in three dimensions and in the full second-order dynamics. At the same time, the numerical study makes clear that the degree of coherence achieved by $T_{\rm osc}$ depends on the competition between gradient-driven ordering, Hubble damping, and the turn-on of the axion potential.

\section{Implications and outlook}

In this work we have proposed a dynamical reinterpretation of axion misalignment in terms of collective phase ordering.
By making explicit the parallel between axion field dynamics and synchronization phenomena, we have shown that a macroscopic axion phase can emerge dynamically from initially incoherent configurations through gradient-driven ordering in an expanding background.

Within this framework, the conventional misalignment angle is not a fundamental initial datum, but an emergent collective variable.
A single macroscopic angle only becomes well defined once phase coherence has developed, i.e. when the order parameter $R(T)$ becomes nonzero.
The collective phase $\Psi(T)$ is then dynamically selected, and the usual identification of the misalignment angle with $\Psi(T_{\rm osc})$ follows naturally.
This emphasizes that both \emph{when} a misalignment angle exists and \emph{which} value it takes are determined by the preceding dynamics.

This perspective also offers a new interpretation of anthropic arguments based on small initial angles.
In the present language, small effective misalignment corresponds to inefficient phase ordering rather than to finely tuned initial conditions.
The apparent need for anthropic selection may therefore be reinterpreted as a statement about the dynamical history of phase coherence, rather than as a requirement on fundamental initial conditions.

The synchronization viewpoint provides a unifying way to understand the standard distinction between Peccei--Quinn symmetry breaking before or after inflation.
Efficient phase ordering over superhorizon scales reproduces the familiar pre-inflationary scenario with a homogeneous axion field, while limited coherence or disruption by topological defects corresponds to the post-inflationary case.
In this sense, the usual taxonomy can be rephrased in terms of the efficiency, scale, and topology of phase ordering dynamics.
From this perspective, axion misalignment is not an arbitrary input, but the outcome of a collective dynamical process in the early Universe.

More broadly, this framework suggests a reorganization of axion initial-condition sensitivity.
Isocurvature fluctuations may be reinterpreted as fluctuations of the collective variables $\Psi$ and $R$, rather than as fluctuations of a fundamental microscopic angle.
From this perspective, the apparent fine tuning of axion initial conditions reflects the dynamics of collective ordering rather than special microscopic choices.

The present three-dimensional study should still be viewed as a robustness analysis rather than a precision cosmological simulation. In particular, a more complete treatment of axion strings, domain-wall dynamics, and quantitatively calibrated cosmological parameters would be needed to assess the detailed efficiency of ordering in realistic post-inflationary scenarios. More generally, the present framework suggests that the relevant question is not simply whether a homogeneous misalignment angle is assumed, but when and to what degree a collective phase becomes dynamically well defined.

More generally, in this framework the axion field may support partially synchronized configurations, in which coherent and incoherent regions coexist.
Such ``chimera'' states are well known in networks of coupled oscillators and spatially extended systems~\cite{KuramotoBattogtokh2002,AbramsStrogatz2004}, and their possible realization in axion cosmology could provide a novel dynamical perspective on inhomogeneous coherence, defect formation, and residual fluctuations.

\begin{acknowledgments}
The work
of V.S. is supported by the Spanish grants PID2023-148162NB-C21,
and CEX2023-001292-S.
\end{acknowledgments}

\appendix

\section{Details of the three-dimensional robustness study}
\label{app:simulation}

\subsection{Dynamical setup}

To test whether the emergence of a collective axion phase survives beyond the original toy illustration, we perform numerical simulations of the full second-order lattice dynamics in three spatial dimensions. The purpose of this appendix is not to claim a precision cosmological prediction, but to document the robustness study summarized in the main text.

We consider a three-dimensional periodic comoving lattice of size $L^3$, with compact phase variable $\theta_{\mathbf n}(t)\in(-\pi,\pi]$ defined at each lattice site. The dynamics is given by Eq.~\eqref{eq:eom_3d_num}. The temperature-dependent axion mass is modeled as a smooth power-law turn-on,
\begin{equation}
m_a(T) =
\begin{cases}
m_0 \left(\dfrac{T}{T_0}\right)^{-n}, & T \ge T_0, \\
m_0, & T < T_0,
\end{cases}
\end{equation}
with representative parameters $T_0=0.15~\mathrm{GeV}$ and $n=4$. The results quoted in the main text correspond to exploratory three-dimensional runs in which we vary both the lattice size and the effective ordering-strength parameter $\Delta x$.

\subsection{Initial conditions and observables}

At the initial temperature, each lattice site is assigned an independent random phase $\theta_{\mathbf n}\in(-\pi,\pi]$, drawn from a uniform distribution. This corresponds to a maximally incoherent state with vanishing expectation value of the complex order parameter up to finite-volume fluctuations. Time evolution is implemented using logarithmic temperature steps, corresponding to uniform fractions of a Hubble time.

The main collective observable is the order parameter
\begin{equation}
R(T)e^{i\Psi(T)}=\frac{1}{N}\sum_{\mathbf n}e^{i\theta_{\mathbf n}(T)},
\end{equation}
where $N=L^3$ is the total number of lattice sites. We also define a practical coherence-onset proxy $u_{\rm coh}$ through the condition $R(u_{\rm coh})=0.5$, where $u\equiv \ln(T_i/T)$ is the cooling variable used in the numerical analysis. Oscillation onset is defined conventionally by $m_a(T_{\rm osc})=3H(T_{\rm osc})$.

\subsection{Interpretation of the numerical study}

The three-dimensional runs summarized in the main text show three robust features. First, the order parameter grows from near-zero values to values close to unity in the baseline full-dynamics runs. Second, this qualitative behavior persists when the effective ordering strength is weakened, although the onset of coherence is delayed and the value of $R(T_{\rm osc})$ decreases. Third, over the lattice sizes explored, the mechanism is stable within the numerical spread of the runs.

We emphasize again that these simulations should be interpreted as a three-dimensional robustness analysis of the qualitative mechanism proposed in this work. They do not yet include a full treatment of strings, domain walls, or quantitatively calibrated cosmological parameters. Their role is instead to show that the emergence of a collective axion phase is not merely an artifact of a one-dimensional overdamped toy model.
\begin{figure}[ht!]
    \centering
    \includegraphics[width=0.7\linewidth]{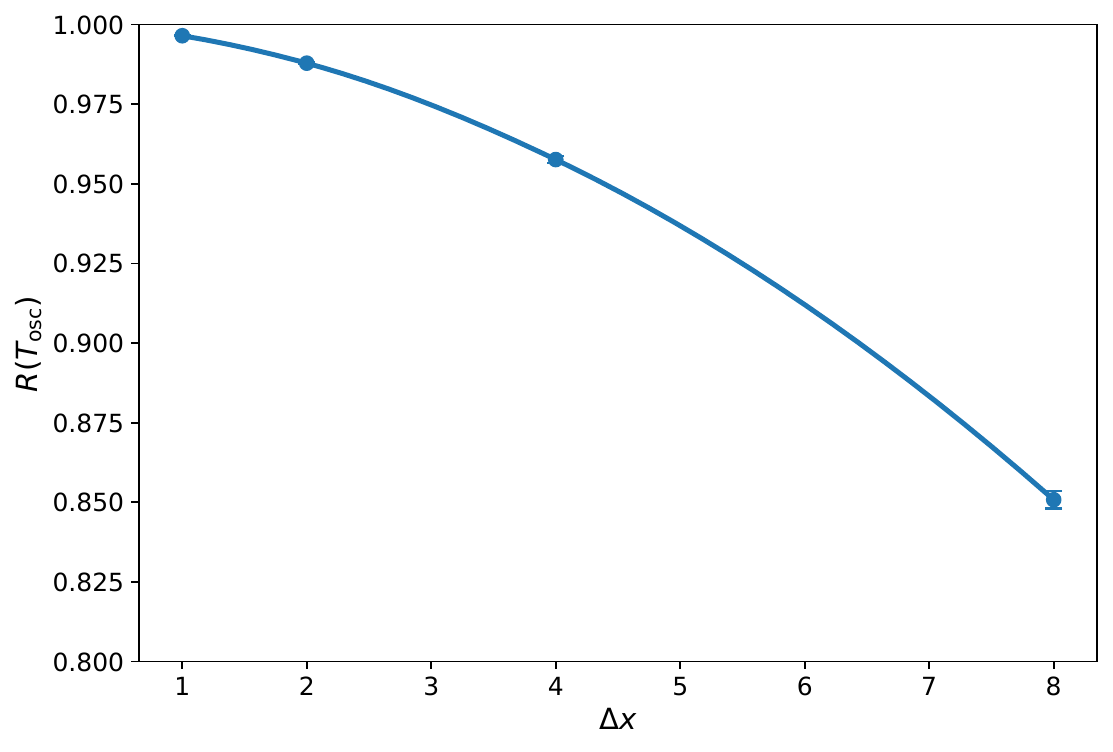}
  \caption{Order parameter at oscillation onset, $R(T_{\rm osc})$, as a function of the ordering-strength parameter $\Delta x$ in the three-dimensional full-dynamics study. The points show the central values extracted from the numerical runs, and the curve is a smooth interpolation included to guide the eye. Over the range explored, increasing $\Delta x$ weakens gradient-driven ordering and lowers the degree of coherence achieved by the time the oscillation condition is reached.}

    \label{fig:3d_dx_scan}
\end{figure}

\bibliographystyle{apsrev4-2}
\bibliography{refs}

@article{PecceiQuinn1977,
  author       = {Peccei, R. D. and Quinn, Helen R.},
  title        = {CP Conservation in the Presence of Instantons},
  journal      = {Phys. Rev. Lett.},
  volume       = {38},
  pages        = {1440--1443},
  year         = {1977},
  doi          = {10.1103/PhysRevLett.38.1440}
}

@article{Beck2013,
  author = {Beck, Christian},
  title = {Axion mass estimate from Josephson junction experiments},
  journal = {Physical Review Letters},
  volume = {111},
  pages = {231801},
  year = {2013}
}

@article{ChadhaDay2014,
  author = {Chadha-Day, F. and Ellis, J. and Marsh, D. J. E.},
  title = {Axion cosmology},
  journal = {Science China Physics, Mechanics \& Astronomy},
  volume = {57},
  pages = {2014},
  year = {2014},
  eprint = {1307.3239},
  archivePrefix = {arXiv},
  primaryClass = {astro-ph.CO}
}

@article{Weinberg1978,
  author       = {Weinberg, Steven},
  title        = {A New Light Boson?},
  journal      = {Phys. Rev. Lett.},
  volume       = {40},
  pages        = {223--226},
  year         = {1978},
  doi          = {10.1103/PhysRevLett.40.223}
}

@article{Wilczek1978,
  author       = {Wilczek, Frank},
  title        = {Problem of Strong $P$ and $T$ Invariance in the Presence of Instantons},
  journal      = {Phys. Rev. Lett.},
  volume       = {40},
  pages        = {279--282},
  year         = {1978},
  doi          = {10.1103/PhysRevLett.40.279}
}

@article{PreskillWiseWilczek1983,
  author       = {Preskill, John and Wise, Mark B. and Wilczek, Frank},
  title        = {Cosmology of the Invisible Axion},
  journal      = {Phys. Lett. B},
  volume       = {120},
  pages        = {127--132},
  year         = {1983},
  doi          = {10.1016/0370-2693(83)90637-8}
}

@article{AbbottSikivie1983,
  author       = {Abbott, L. F. and Sikivie, Pierre},
  title        = {A Cosmological Bound on the Invisible Axion},
  journal      = {Phys. Lett. B},
  volume       = {120},
  pages        = {133--136},
  year         = {1983},
  doi          = {10.1016/0370-2693(83)90638-X}
}

@article{DineFischler1983,
  author       = {Dine, Michael and Fischler, Willy},
  title        = {The Not-So-Harmless Axion},
  journal      = {Phys. Lett. B},
  volume       = {120},
  pages        = {137--141},
  year         = {1983},
  doi          = {10.1016/0370-2693(83)90639-1}
}

@article{Turner1986,
  author       = {Turner, Michael S.},
  title        = {Cosmic and Local Mass Density of Invisible Axions},
  journal      = {Phys. Rev. D},
  volume       = {33},
  pages        = {889--896},
  year         = {1986},
  doi          = {10.1103/PhysRevD.33.889}
}

@article{Sikivie2008,
  author       = {Sikivie, Pierre},
  title        = {Axion Cosmology},
  journal      = {Lect. Notes Phys.},
  volume       = {741},
  pages        = {19--50},
  year         = {2008},
  doi          = {10.1007/978-3-540-73518-2_2}
}

@article{Marsh2016,
  author       = {Marsh, David J. E.},
  title        = {Axion Cosmology},
  journal      = {Phys. Rept.},
  volume       = {643},
  pages        = {1--79},
  year         = {2016},
  doi          = {10.1016/j.physrep.2016.06.005}
}

@article{Kibble1976,
  author       = {Kibble, T. W. B.},
  title        = {Topology of Cosmic Domains and Strings},
  journal      = {J. Phys. A},
  volume       = {9},
  pages        = {1387--1398},
  year         = {1976},
  doi          = {10.1088/0305-4470/9/8/029}
}

@article{Bray1994,
  author       = {Bray, Alan J.},
  title        = {Theory of Phase-Ordering Kinetics},
  journal      = {Adv. Phys.},
  volume       = {43},
  pages        = {357--459},
  year         = {1994},
  doi          = {10.1080/00018739400101505}
}

@article{Acebron2005,
  author       = {Acebr{\'o}n, Juan A. and Bonilla, L. L. and Vicente, Conrad J. P. and Ritort, F. and Spigler, R.},
  title        = {The Kuramoto Model: A Simple Paradigm for Synchronization Phenomena},
  journal      = {Rev. Mod. Phys.},
  volume       = {77},
  pages        = {137--185},
  year         = {2005},
  doi          = {10.1103/RevModPhys.77.137}
}

@article{Kuramoto1975,
  author       = {Kuramoto, Yoshiki},
  title        = {Self-entrainment of a population of coupled non-linear oscillators},
  journal      = {Prog. Theor. Phys.},
  volume       = {54},
  pages        = {687--699},
  year         = {1975},
  doi          = {10.1143/PTP.54.687}
}

@article{KuramotoBattogtokh2002,
  author = {Kuramoto, Y. and Battogtokh, D.},
  title = {Coexistence of coherence and incoherence in nonlocally coupled phase oscillators},
  journal = {Nonlinear Phenomena in Complex Systems},
  volume = {5},
  pages = {380--385},
  year = {2002}
}

@article{AbramsStrogatz2004,
  author = {Abrams, D. M. and Strogatz, S. H.},
  title = {Chimera states for coupled oscillators},
  journal = {Physical Review Letters},
  volume = {93},
  pages = {174102},
  year = {2004}
}

@book{Kuramoto1984,
  author       = {Kuramoto, Yoshiki},
  title        = {Chemical Oscillations, Waves, and Turbulence},
  publisher    = {Springer},
  address      = {Berlin},
  year         = {1984},
  isbn         = {978-3540134013}
}

\end{document}